\documentclass[preprint2]{aastex}

\begin{document}

\title{Schwarzschild black hole in dark energy background}

\author{Ngangbam Ishwarchandra\dag1, \quad Ng. Ibohal\S \quad and K. Yugindro Singh\dag}
\affil{\dag Department of Physics, Manipur University, Imphal - 795003, Manipur, INDIA \\
\S Department of Mathematics, Manipur University, Imphal - 795003, Manipur, INDIA}
\email{E-mail: \dag1 ngishwarchandra@manipuruniv.ac.in, \dag1 ngishwarchandra@gmail.com, \dag yugindro361@gmail.com\\ \S ngibohal@iucaa.ernet.in and \S ngibohal@manipuruniv.ac.in}

\begin{abstract}
In this paper we present an exact solution of Einstein's field equations describing  the Schwarzschild black hole in dark energy background. It is also regarded as an embedded solution that the Schwarzschild black hole is embedded into the dark energy space producing Schwarzschild-dark energy black  hole. It is found that the space-time geometry of Schwarzschild-dark energy solution is non-vacuum Petrov type $D$ in the  classification of space-times. We study the energy conditions (like weak, strong and dominant conditions) for the energy-momentum tensor of the Schwarzschild-dark energy solution. We also find that the energy-momentum tensor of the Schwarzschild-dark energy solution violates the strong energy condition due to the negative pressure leading to a repulsive gravitational force of the matter field in the space-time. It is shown that the time-like vector field for an observer in the Schwarzschild-dark energy space is expanding, accelerating, shearing and non-rotating.   We investigate the surface gravity and the area of the horizons for the Schwarzschild-dark energy black hole.\\

\end{abstract}
\keywords{Schwarzschild solution; dark energy; exact solutions; energy conditions; surface gravity.}
\section{Introduction}
\setcounter{equation}{0}
\renewcommand{\theequation}{1.\arabic{equation}}
In general relativity the Schwarzschild solution is regarded as a black hole in an asymptotically flat space. The Schwarzschild-de Sitter solution is interpreted as a black hole in an asymptotically de Sitter space with non-zero cosmological constant $\Lambda$ \citep{gih77}. The Schwarzschild-de Sitter solution is also considered as an embedded black hole that the Schwarzschild solution is embedded into the de Sitter space to produce the Schwarzschild-de Sitter black hole \citep{cai98}. Here we are looking for an exact solution to describe the Schwarzschild black hole in an asymptotically dark energy space as Schwarzschild-dark energy black hole.

Here we mean the dark energy solution as the one possessing a non-perfect fluid energy-momentum tensor having the equation of state parameter $\omega=p/\rho=-1/2$ with negative pressure presented in \citet{iis11}. For deriving the Schwarzschild-dark energy solution we adopt the mass function expressed in a power series expansion of the radial coordinate in \citet{wanwu99} as
\begin{equation}
\hat{M}(u,r)= \sum_{n=-\infty}^{+\infty} q_n(u)\,r^n,
\end{equation}
where $q_n(u)$ are arbitrary functions of retarded time coordinate $u$. The important role of the Wang-Wu mass function $M(u,r)$ in generating new exact solutions of Einstein's field equations are explained in \citet{iis11}. The mass function is being utilized in deriving {\sl non-rotating} embedded Vaidya solution  into other spaces by choosing the function $q_n(u)$ corresponding to the number $n$ \citep{wanwu99}. Further utilizations of the mass function have been extended in rotating system and found the role of the number $n$ in generating rotating embedded solutions of field equations \citep{ibo05}. Here we shall consider the values of the index number $n$ as $n=0$ and $n=2$. That the value $n = 0$ corresponds to the Schwarzschild solution and $n = 2$ for the dark energy solution possessing the equation of state parameter $\omega=-1/2$.
These values of $n$ will conveniently combine in order to obtain embedded Schwarzschild-dark energy black hole. The resulting solution will be an extension of the work presented in \citet{iis11} with $n=2$. The difference between the works of this paper and that of \citet{iis11} may be seen in the following sections. It is better to recall that the Schwarzschild-de Sitter black hole is the combination of two solutions corresponding to the index number $n=0$ (Schwarzschild) and $n=3$ (de Sitter), different from the Schwarzschild-dark energy solution with $n=0$ and $n=2$.

Sections 2.2 deals with the derivation of the embedded Schwarzschild-dark energy black hole of Einstein's equation. We find that the masses of the  solution proposed here describe the gravitational fields of the space-time geometry and also determine the matter distributions with negative pressure, whose energy equation of state has the value $-1/2$. However, the energy-momentum tensor of the matter distribution of Schwarzschild-dark energy solution with negative pressure does not describe a perfect fluid. The non-perfect fluid distribution is in agreement with the Islam's remark (1985) that ``it is not necessarily true that the field is that of a star made of perfect fluid''. We also find that each solution has a coordinate singularity with  horizon. Consequently we discuss the area, entropy and surface gravity at the horizons for the solution. The existence of the horizons discussed here is also in accord with the cosmological horizon \citep{gih77} of de Sitter space with constant $\Lambda$, which is usually considered to be a common  candidate of dark energy with the equation of state parameter $w = -1$. The paper is concluded in Section 3 with reasonable remarks and evolution of the solutions with the physical interpretation. It is interesting to concise the results of the paper in the forms of the following theorems:

\newtheorem{theorem}{Theorem}
\begin{theorem}
The embedded Schwarzschild-dark energy solution admitting an energy-momentum tensor having negative pressure with equation of state parameter $w=-1/2$, is a non-vacuum Petrov type  $D$ space-time.
\end{theorem}
\begin{theorem}
The energy-momentum tensor of the matter distribution in the embedded Schwarzschild-dark energy solution violates the strong energy condition leading to a repulsive gravitational force in the geometry.
\end{theorem}
\begin{theorem}
The time-like vector fields of the matter distributions in the embedded Schwarzschild-dark energy space-time is expanding, accelerating and shearing with zero-twist.
\end{theorem}
\begin{theorem}
The surface gravity at the horizon of extreme Schwarzschild-dark energy space-time is vanished.
\end{theorem}
\begin{theorem}
The Schwarzschild-dark energy solution can be expressed in the Kerr-Schild ansatze on different backgrounds.
\end{theorem}
Theorem 1 shows the physical interpretation of the solution that all  components of the Weyl tensors of the Schwarzschild-dark energy space-time vanish except $\psi_2$ indicating Petrov D space-time. The energy-momentum tensor associated with the solution admits dark energy having the negative pressure and the energy equation of state parameter $w=-1/2$. It is also found that the mass of the solution not only describe the gravitational field in the space-time geometry, but also measure the energy density and the negative pressure in the energy-momentum tensor indicating the {\it non-vacuum} status of each solution. It is the assertion of General Relativity that ``the space-time geometry is influenced by the matter distribution'' \citep{wal84}. The violation of the strong energy condition due to the negative pressure is shown in Theorem 2 leading to a repulsive gravitational field of the space-time geometry. Theorem 3 shows the physical interpretation of time-like vector fields of the matter distributions. Theorem 4 indicates the non-existence of surface gravity on the horizon of the extreme Schwarzschild-dark energy black hole. It is to mention that the non-embedded Schwarzschild solution represents a {\it vacuum}, non-conformally flat space-time showing the difference from the {\it non-vacuum}, non-conformally flat Schwarzschild-dark energy solution discussed in this paper.
\section{Schwarzschild-dark energy black hole}
\setcounter{equation}{0}
\renewcommand{\theequation}{2.\arabic{equation}}

In this section we will derive an embedded Schwarzschild-dark energy solution of Einstein's field equations. This solution will describe Schwarzschild black hole in asymptotically dark energy background as the Schwarzschild-de Sitter black hole is regarded as  Schwarzschild black hole in asymptotically de Sitter space \citep{gih77}. For deriving an embedded Schwarzschild-dark energy solution, we consider the mass function in the expansion series (1.1) by choosing the Wang-Wu function $q_{n}(u)$ as
\begin{eqnarray}
\begin{array}{cc}
q_n(u)=&\left\{\begin{array}{ll}
M, &{\rm when}\;\;n=0\\
m, &{\rm when}\;\;n=2\\
0, &{\rm when }\;\;n\neq 0,2,
\end{array}\right.
\end{array}
\end{eqnarray}
such that the mass function (1.1) takes the form
\begin{equation}
M(u,r)\equiv \sum_{n=-\infty}^{+\infty} q_n(u)\,r^n =M+mr^2.
\end{equation}
where $m$ is constant which will later be considered as the mass of the dark energy.
Then using this mass function $q_{n}(u)$ in general canonical metric in Eddington-Bekenstein coordinate system
\begin{eqnarray*}
ds^2=\Big\{1-\frac{2M(u,r)}{r}\Big\} du^2+2du\,dr-r^2d\Omega^2,
\end{eqnarray*}
with $d\Omega^2=d\theta^2+{\rm sin}^2\theta\,d\phi^2$,
we find a line element in the null coordinates system $(u,r,\theta,\phi)$
\begin{eqnarray}
ds^2&=&\Big\{1-{2\over r}(M+mr^2)\, \Big\}\,du^2 +2du\,dr\cr &&
-r^2d\theta^2-{r^2
\rm sin}^2\theta\,d\phi^2,
\end{eqnarray}
where the constant $m$ is regarded as the mass of the dark energy and is non-zero for the existence of the dark energy distribution in the universe. M is the mass of Schwarzschild solution. The line-element will reduce to that of Schwarzschild black hole when $m=0$ with singularity at $r=2M$, and also it will be that of dark energy when $M=0$ having singularity at $r=(2m)^{-1}$. This line-element (2.3) will have horizons as the roots of the equation  $\Delta\equiv r^2-2r(M-r^2m)=0$, that is, the roots are $r_\pm =(1/4m)\{1\pm\sqrt{1-16mM}\,\}$.

We obtain the null tetrad vectors  for the metric line element as follows
\begin{eqnarray}
&&\ell_a=\delta^1_a, \cr
&&n_a=\frac{1}{2}\Big\{1-2r^{-1}(M+mr^2)\Big\}\,\delta^1_a+ \delta^2_a, \cr
&&m_a=-{r\over\surd 2}\,\Big\{\delta^3_a +i\,{\rm
sin}\,\theta\,\delta^4_a\Big\},\\
&&m_a=-{r\over\surd 2}\,\Big\{\delta^3_a -i\,{\rm
sin}\,\theta\,\delta^4_a\Big\}, \nonumber
\end{eqnarray}
where $\ell_a$,\, $n_a$
are real null vectors and $m_a$ is complex having its conjugate $\bar{m}_a$ with the normalization conditions $\ell_an^a= 1 = -m_a\bar{m}^a$ and other inner products are zero.
The Newman-Penrose spin coefficients, Ricci scalars of the embedded Schwarzschild-dark energy black hole are obtained as follows:
\begin{eqnarray}
&&\kappa=\epsilon=\sigma=\nu=\lambda=\pi=\tau=0, \cr
&& \rho=-{1\over r}, \quad
\beta=-\alpha={1\over {2\surd 2r}}\,{\rm cot}\theta, \cr
&&\mu=-{1\over 2\,r}\Big\{1-{2M\over r}-2rm\Big\}, \cr
&&\gamma={1\over 2\,r^2}\,\Big\{M -mr^2\Big\}, \\
&&\phi_{11}=\frac{1}{2r}m,\cr
&&\Lambda^* = \frac{1}{2r}m,
\end{eqnarray}
From the Einstein's field equation the above space-time (2.3)
possesses an energy-momentum tensor (stress-energy tensor) describing dark energy distribution in the gravitational field as
\begin{eqnarray}
T_{ab} &=& 2\,\rho\,\ell_{(a}\,n_{b)}
+2\,p\,m_{(a}\bar{m}_{b)},
\end{eqnarray}
where the quantities are found as
\begin{eqnarray}
&& \rho = {4\over Kr}m, \quad p = -{2\over Kr}m.
\end{eqnarray}
These $\rho$ and $p$ will be interpreted as the density and pressure respectively for dark energy with the universal constant $K=8\pi G/c^4$. The equation (2.8) indicates that the contribution of the gravitational field  to $T_{ab}$ is measured directly by mass $m$ with negative pressure $p$. The energy-momentum tensor (2.7) is calculated from Einstein's equations $R_{ab} - (1/2)Rg_{ab} = -KT_{ab}$ of gravitational field for the space-time metric (2.3) with the relation $K\rho = 2\,\phi_{11} + 6\,\Lambda^*$ and $Kp = 2\,\phi_{11} - 6\,\Lambda^*$, where $\phi_{11}$ and $\Lambda^*$ are Ricci scalars given in (2.6). Here from (2.8) we observe the key role of the mass $m$ that it not only describes the curvature of the space-time in (2.3) but also distributes the matter (2.7) content in the space-time.
Then, the equation of state for the solution (2.3) is found from (2.8) as
\begin{eqnarray}
\omega=\frac{p}{\rho}=-{1\over2}.
\end{eqnarray}
 This energy-momentum tensor (2.7) satisfies the energy conservation law given in Ibohal (2009) in Newman-Penrose formalism \citep{newpen62}.
\begin{eqnarray}
T^{ab}_{\;\;\;\,;b}=0,
\end{eqnarray}
which shows the fact that the metric of the line element (2.3) describing dark energy is a solution of Einstein's field equations.
The component of energy-momentum tensor may be written for future use as:
\begin{eqnarray}
T_{u}^{u}=T_{r}^{r}=\rho, \quad
T_{\theta}^{\theta}=T_{\phi}^{\phi}=-p.
\end{eqnarray}
We find the trace of the energy momentum tensor
$T_{ab}$ (2.7) as follows
\begin{equation}
T=2(\rho-p)=\frac{12}{Kr}m.
\end{equation}
Here $\rho - p$ must be always greater than zero for the existence of the Schwarzschild-dark energy solution $(2.3)$ with $m \neq 0$, (if $\rho=p$ implies that $m$ will vanish). This negative pressure strongly supports the interpretation of the line element (2.3) of a Schwarzschild-dark energy solution of Einstein's field equations with the energy-momentum tensor (2.7).

It is to emphasize that the energy momentum
tensor (2.7) does not describe a perfect fluid, i.e. for a
non-rotating perfect fluid $T^{(\rm {pf})}_{ab} =
(\rho+p)u_a\,u_b-p\,g_{ab}$ with unit time-like vector $u_a$ and
trace $T^{(\rm {pf})}=\rho-3p$, which is different from the one
given in (2.12).

Then the energy momentum tensor (2.7) can be written as
\begin{eqnarray}
T_{ab}=(\rho+p)(u_{a}u_{b}-v_{a}v_{b})-pg_{ab}.
\end{eqnarray}
where $u_{a}=\frac{1}{\sqrt{2}}(l_a+n_a)$ and $v_{a}=\frac{1}{\sqrt{2}}(l_a-n_a)$.
We observe from the energy density and the pressure (2.8) that the energy-momentum tensor obeys the {\it weak energy condition} $T_{ab}U^aU^b \geq 0$ for any future directed time-like vector $U_{a}=\hat{\alpha}u_{a}+\hat{\beta}v_{a}+\hat{\gamma}w_{a}+\hat{\delta}z_{a}$ \citep{{csr03},{iis11}}
\begin{eqnarray}
\rho\geq0, \quad \rho+p\geq0.
\end{eqnarray}
and the {\it dominant energy condition} that $T_{ab}U^b$ should be a future directed non-space like  vector field
\begin{eqnarray}
{\rho}^2\geq0, \quad  {\rho}^2-p^2\geq0.
\end{eqnarray}
However, $T_{ab}$ violates the {\it strong energy condition} $R_{ab}\,U^aU^b\geq0$
\begin{eqnarray}
p\geq0,\quad \rho+p\geq0.
\end{eqnarray}
This violation of the strong energy condition is due to the negative pressure (2.8), and implies that the gravitational force of the dark energy is repulsive which may cause the acceleration of the model, like the cosmological constant leads to the acceleration of the expansion of the Universe.

We find the Weyl scalars of the Schwarzschild-dark energy black hole describing gravitational field
\begin{eqnarray}
\psi_0 =\psi_1 =\psi_3 =\psi_4 =0,\quad
\psi_2 =-\frac{1}{r^3}M.
\end{eqnarray}
The non-zero Weyl scalar $\psi_2$ indicates that the space-time of the embedded solution is Petrov type $D$ in the classification of space-time. It is also observed that the mass $m$ of the dark energy  does not appear in  (2.17) showing the intrinsic property  of the conformally flatness of the dark energy even embedded into the Schwarzschild black hole. From (2.8) and (2.17), it concludes the proof of the Theorem 1 that the Schwarzschild-dark energy black hole is a non-vacuum Petrov type $D$ space-time possessing the equation of state parameter $w=-1/2$.

The curvature invariant for the Schwarzschild-dark energy model (2.3) is found as
\begin{eqnarray}
R_{abcd}R^{abcd}={48\over r^6}M^2+{32\over r^2}m^2.
\end{eqnarray}
This invariant diverges only at the origin $r=0$ showing that the origin $r=0$ is a physical singularity. This indicates that the singularity of the solution (2.3) is caused due to the coordinate system, such like in Schwarzschild solution when $m=0$.

\subsection{Kerr-Schild ansatze of Schwarzschild-dark energy metric}
Here the important characteristic feature of embedded solution will be discussed that we cannot physically predict which space of the two solutions started first to embed into another \citep{ibo05}. We shall clarify the nature of the embedded solution in the form of Kerr-Schild ansatze in different backgrounds.  For this we express the Schwarzschild-dark energy metric in
Kerr-Schild ansatz on the Schwarzschild background first as
\texttt{}\begin{equation}
g_{ab}^{(\rm SchDE)}=g_{ab}^{(\rm Sch)} +2Q(r)\ell_a\ell_b,
\end{equation}
where $Q(r) = - mr$. Here,
$g_{ab}^{(\rm Sch)}$ is the Schwarzs-child metric and $\ell_a$ is  geodesic, shear free, expanding and zero twist null vector for
both $g_{ab}^{(\rm Sch)}$ as well as $g_{ab}^{(\rm SchDE)}$. The above
Kerr-Schild form can also be recast on the dark energy metric  background
as
\begin{equation}
g_{ab}^{\rm (SchDE)}=g_{ab}^{\rm (DE)} +2Q(r)\ell_a\ell_b,
\end{equation}
where $Q(r) =-Mr^{-1}+e^2r^{-2}/2$. These two
Kerr-Schild forms (2.19) and (2.20) show the fact that the embedded Schwarzschild-dark energy space-time (2.3) with the mass $m$ of the dark energy is a solution of Einstein's field equations. They establish the structure of embedded black hole that  either ``the Schwarzschild black hole is embedded into the dark energy space to obtain Schwarzschild-dark energy black hole" or ``the dark energy space-time is embedded into the Schwarzschild black hole to obtain the dark energy-Schwarzschild black hole'' -- both  nomenclatures (Schwarzschild-dark energy and dark energy-Schwarzschild) possess geometrically the same physical structure. This is the important characteristic feature of embedded solutions that it is hard physically to predict which space of the two started first to embed into another (Ibohal, 2005). Here we come to the conclusion of the proof of theorem 5 stated above.

\subsection{Raychaudhuri equation of Schwarzschild black hole in dark energy}
It is important to study the nature of the time-like vector $u^a$ appeared in the energy-momentum tensor (2.13) for the dark energy model (2.3). In fact it describes the physical properties of the matter whether the matter is expanding ($\Theta=u^a_{\:\,;a} \neq 0$), accelerating ($\dot{u}_a=u_{a;b}u^b\neq 0$) shearing $\sigma_{ab}\neq0$ or non-rotating ($w_{ab}=0$). We shall investigate the rate of expansion from the Raychaudhuri equation, such that we can understand how the negative pressure of the dark energy affects the expansion rate of the solution.
We write the explicit form of the time-like vector as
\begin{eqnarray}
u_a&=&\frac{1}{\sqrt{2}}(\ell_a+n_a)\cr
&=&\frac{1}{\surd{2}}\Big[1+\frac{1}{2}\{1-2r(M+r^2m)\}\Big]\delta^1_a\cr &&
+\frac{1}{\surd 2}\delta^2_a.
\end{eqnarray}
For this purpose we find the covariant derivative of the time-like vector $u^a$ in terms of null tetrad vectors as follows
\begin{eqnarray}
u_{a;b}&=&\frac{1}{\sqrt{2}}\Big[{1\over r^2}(M-mr^2)(\ell_a\ell_b-n_a\ell_b)\cr &&-\frac{1}{2r}\Big\{1+\frac{2M}{r}+2rm\Big\}\cr &&\times(m_a\bar{m}_b+\bar{m}_a m_b)\Big],
\end{eqnarray}
where $m$ is the mass of the dark energy and $M$ that of Schwarzschild black hole.  This expression of $u_{a;b}$ is convenient to obtain the expansion scalar $\Theta=u^a_{\:\,;a}$ and acceleration vector $\dot{u}_a=u_{a;b}u^b$ as follows
\label{eq:whole}
\begin{eqnarray}
&&\Theta = \frac{1}{\surd 2\,r^2}\Big(r+M+3r^2m\Big) \\ \label{subeq:1}
&&\dot{u}_a=-\frac{1}{2r^2}\Big\{M-mr^2\Big\}(\ell_a - n_a) \\ \label{subeq:2}
&&\dot{u}^a_{\:\,;a}=\frac{1}{2r^4}\,M^2-\frac{3m}{2r}(1-rm). \label{subeq:3}
\end{eqnarray}
We find that for the non-rotating solution (2.3), the vorticity tensor $w_{ab}$ is vanished,
and however, the shear tensor $\sigma_{ab}=\sigma_{(ab)}$ exists as
\begin{equation}
\sigma_{ab}=\frac{1}{3\surd 2 r^2}(r+4M)(v_av_b-m_{(a}\bar{m}_{b)}),
\end{equation}
which is orthogonal to $u^a$ ({\it i.e.,} $\sigma_{ab}u^b=0$). It is found that the mass $m$ of the dark energy does not explicitly involve in the expression  of $\sigma_{ab}$ but its involvement can be seen in the null vector $n_a$ (2.4) as $v_a=(1/\surd 2)(\ell_a-n_a)$. However, the mass $m$ directly determines the expansion as well as the acceleration of the model as seen in (2.23). We find from (2.23) that the solution discussed here follows the non-geodesic path of the time-like vector ($u_{a:b}u^b\neq 0$). This establishes the key result of the expansion of the dark energy with acceleration. The vanishing of the vorticity tensor $w_{ab}=0$ can be interpreted physically  as saying that the dark energy is non-rotating ({\it twist-free}) as mentioned earlier. From the above it follows the proof of the Theorem 3 cited in the introduction.

Now let us observe the consequences of the Raychaudhuri equation for the solution (2.3). The Raychaudhuri equation is given by
\begin{equation}
\dot{\Theta}=\dot{u}^a_{\:\,;a}+2(w^2-\sigma^2)-\frac{1}{3}\theta^2 +R_{ab}u^a u^b
\end{equation}
and $\dot{\theta}=\theta_{;a}u^{a}$ where the shear and vorticity magnitudes are $2\sigma^2=\sigma_{ab}\sigma^{ab}$ and $2w=w_{ab}w^{ab}$; $R_{ab}$ is the Ricci tensor of the space-time metric $g_{ab}$ of (2.3). Then the Raychaudhuri equation for our  twist-free time-like vector
is found as follows
\begin{equation}
\dot{\Theta}=-\frac{1}{4r^2}(1+2rm)-\frac{M}{r^4}(r+M+r^2m),
\end{equation}
where the negative pressure $p$ given in (2.8) of the dark energy is taken care in Ricci tensor $R_{ab}$.
\subsection{Surface Gravity of Schwarzschild-dark energy black hole}
Here we shall investigate the surface gravity of Schwarzschild-dark energy black hole on the horizon. The equation $\Delta=0$ of the embedded Schwarzschild-dark energy black hole (2.3) will give the horizons i.e, the polynomial
\begin{equation}
\Delta\equiv 1-\frac{2}{r}(M+r^2m)=0.
\end{equation}
has two roots
\begin{eqnarray}
r_\pm=\frac{1}{4m}\{1\pm\sqrt{1-16mM}\,\}.
\end{eqnarray}
The root $r_{+}$ corresponds the event horizon and the root $r_{-}$ the apparent horizon for the Schwarzschild-dark energy black hole (2.3). Then we can also obtain the areas of the horizons for the embedded dark energy black hole as
\begin{eqnarray}
A&=&4\pi r^2_{\pm} \cr
&=&\frac{\pi}{4m^2}\{1\pm\sqrt{1-16mM}\,\}^2.
\end{eqnarray}
Now from the entropy-area formula $S=A/4$ for black holes \citep{gih77}, we have the entropies of the horizons
\begin{eqnarray}
S&=&\frac{A}{4} \cr
&=&\frac{\pi}{16m^2}\{1\pm\sqrt{1-16mM}\,\}^2.
\end{eqnarray}
The surface gravity $\kappa$ of the Schwarzschild-dark energy black hole on the horizons $r=r_{\pm}$ can be obtained as follows
\begin{eqnarray}
\kappa=\frac{1}{2}\Delta'={1\over r^2}\,\Big\{M-mr^2\Big\}\Big|_{r=r_{\pm}}.
\end{eqnarray}
It is observed that when $1-16mM=0$, that is $M=(1/16m)$, the Schwarzschild-dark energy solution may become the extreme black hole with the horizon $r_{+}=r_{-}=(1/4m)$. Then the surface gravity of the Schwarzschild-dark energy black hole will be vanished $\kappa=0$ on the horizon $r=r_{\pm}$. This completes the proof of the Theorem 4 that ``the surface gravity at the horizon of extreme Schwarzschild-dark energy space-time is vanished.''

It is to note that when the dark energy mass sets to zero $m=0$, the surface gravity (2.33) of the Schwarzschild-dark energy black hole will reduce to that of Schwarzschild black hole \citep{poi04}
\begin{eqnarray}
\kappa=\frac{1}{4M}.
\end{eqnarray}
on the horizon $r=2M$. The surface gravity plays an important characteristic role that it provides the Hawking temperature $T=\kappa/2\pi$ of the Schwarzschild-dark energy black hole on the horizons $r=r_{\pm}$ (2.30).
\section{Conclusion }
In this paper we develop an exact solution  of Einstein's field equations describing embedded Schwarzschild-dark energy black hole, which is a non-vacuum Petrov type $D$ space-time. The energy-momentum tensor possesses dark energy fluid with the negative pressure and the equation of state parameter $w=-1/2$. The important property of the solution is that the  metric tensor $g_{ab}$ describes both the background space-time structure and the dynamical aspects of the gravitational field in the form of energy-momentum tensor.

That is to mention that the masses of the solutions play the role of both the curvature of the space-time ({\it non-flat}) as well as the source of the energy-momentum tensor with $T_{ab}\neq 0$ ({\it non-vacuum}) measuring the energy density and the negative pressure. This indicates that when we set $m=0$ and $M=0$ of the solution to be zero, the space-times will become the flat Minkowski space with vacuum structure $T_{ab}=0$. In the case of Schwarzschild solution, the mass plays only the role of curvature of the space-time and cannot determine the energy-momentum tensor. That is why the Schwarzschild solution is a curved {\it non-flat, vacuum} space-time with $T_{ab}=0$. Here lies the advantage of the solutions (2.3) as {\it non-flat} and {\it non-vacuum} space-times. It is also to mention that the Schwarzschild-dark energy solution discussed here provides example of non-conformally flat space-times. However, the dark energy solution with $m\neq 0$ and $M=0$ is a conformally flat spacetime, while other examples of conformally flat solutions are the non-rotating de Sitter models with cosmological constant $\Lambda$ \citep{hael73}, cosmological function $\Lambda(u)$ \citep{ibo09} and the Robertson-Walker metric \citep{ste85}.

We find that the time-like vector of the observer is expanding $\Theta \neq 0$ (2.23), accelerating $\dot{u}_a\neq 0$ (2.24) as well as shearing $\sigma_{ab} \neq 0$ (2.26), but non-rotating $w_{ab}=0$. The equation (2.24) $\dot{u}_a=u_{a;b}u^b \neq 0$ means that the stationary observer of the solution does not follow the time-like geodesic path  $u_{a;b}u^b = 0$. We also find that the energy-momentum tensor for the solution violates the strong energy condition. The violation of strong energy condition is due to the negative pressure of the matter field content in the space-time geometry, which can be seen in (2.8) above, and is not an assumption to obtain the solution (like other dark energy models mentioned in Sahni 2004). This violation indicates that the gravitational field of the solution is repulsive (as pointed out in Tipler (1978) leading to the accelerated expansions of the universe. The expansion of the space-time with acceleration is in agreement with the observational data (Permutter, et al. 2000 and Riess, et al. 2000, 2001). It is also noted that the strong energy condition for the solution (2.3) associated with the stress-energy momentum tensor (2.8) is different from that of the perfect fluid ($\rho \geq 0$, $\rho + 3p \geq 0$). This indicates that the strong energy condition is mainly depended upon the internal structure of a particular matter field distribution in the spacetime.

It is emphasized the fact that our approach in the development of the solutions here is necessarily based on the identification of the power index $n=0, 2$ in the Wang-Wu mass function expansion (1.1) without any extra assumption. It is also noted that we do not consider the Friedman-Robertson-Walker metric, filled with perfect fluid, which is assumed to be the standard approach for the investigation of dark energy problem  \citep{{bou08}, {pad03}, {cop06}, {sqg05}, {got07}, {rp88}, {joh02}, {cds98}, {fr04}}. That is why the energy-momentum tensors associated with the solutions (2.3) do not describe a perfect fluid. This fact can be observed from the trace $T=2(\rho-p)$ of the $T_{ab}$ for the solution (2.3). This non-perfect fluid  distribution of the solution is also in accord with Islam's suggestion that it is not necessarily true that stars are made of perfect fluid (Islam 1985).

From the study of the above solutions, we find that the energy densities are only contributed  from the masses of the matter. It is the fact that without the mass of the solution, one cannot measure the energy density and the negative pressure of the energy-momentum tensor in order to obtain the energy equation of state $w=-1/2$. This means that the negative pressure and the energy density associated with the energy-momentum tensor (2.13) are measured by the masses that produce the gravitational field in the space-time geometry of the solution. Hence, we may conclude that the solution (2.3) may explain the essential part of Mach's principle -- ``the matter distribution influences the space-time geometry'' (Wald, 1984). It is emphasized that the equation of state parameters $w=-1/2$ for the matter distribution (2.7) are belonged to the range $-1<w<0$ focussed for the best fit with cosmological observations in \citet{cds98} and references there in.

The existence of horizons is also in accord with the cosmological horizon of de Sitter space with constant $\Lambda$ \citep{gih77}, which is considered to be a common candidate of dark energy with the parameter $w= - 1$ \citep{{bou08}, {pad03}, {cop06}, {sqg05}, {got07}, {rp88}, {joh02}, {cds98}, {fr04}}. This parameter of equation of state is also true for both the cosmological constant $\Lambda$ as well as the cosmological function $\Lambda(u)$ of the {\it rotating} and {\it non-rotating} de Sitter solutions \citep{ibo09}. We hope that the exact solution (2.3) may provide an example of space-times admitting energy-momentum tensor having negative pressure with the equation of state parameter $w=-1/2$ possessing with a time-like vector field.

\end{document}